\documentstyle[12pt]{article}
\begin{document}
\begin{titlepage}
\rightline{SNUTP/96-12}
\def\today{\ifcase\month\or
        January\or February\or March\or April\or May\or June\or
        July\or August\or September\or October\or November\or December\fi,
  \number\year}
\rightline{hep-th/9603028}
\rightline{March, 1996}
\vskip 1cm
\centerline{\Large \bf Integrable structure in  
supersymmetric gauge theories} 
\centerline{\Large \bf with massive hypermultiplets}
\vskip 2cm
\centerline{{\sc Changhyun Ahn} and
{\sc Soonkeon Nam} \footnote{
nam@nms.kyunghee.ac.kr}}
\vskip 1cm
\centerline {{\it Department of Physics and}} 
\centerline {{\it Research Institute for Basic Sciences,}}
\centerline {{\it Kyung Hee University, Seoul 130-701, Korea}}
\vskip 1cm
\centerline{\large \sc Abstract}
\vskip 0.2in
We study the quantum moduli space of vacua of $N=2$ supersymmetric 
$SU(N_c)$ gauge theories coupled to $N_f$ flavors of quarks
in the fundamental representation. We identify the moduli space of the  
$N_c = 3$ and $N_f=2$  massless case with the full spectral curve
obtained from the Lax representation of the Goryachev-Chaplygin top.
For the case with {\it massive} quarks, we present an integrable system 
where the corresponding hyperelliptic curve parametrizing the 
Laurent solution coincides with that of the moduli space
of $N_{c}=3$ with $N_{f}=0, 1, 2$.
We discuss possible generalizations of the integrable
systems relevant to gauge theories with $N_c \neq 3 $ and general $N_f$. 

\end{titlepage}
\newpage

\def\beq{\begin{equation}}
\def\eeq{\end{equation}}
\def\bea{\begin{eqnarray}}
\def\eea{\end{eqnarray}}
\renewcommand{\arraystretch}{1.5}
\def\ba{\begin{array}}
\def\ea{\end{array}}
\def\bce{\begin{center}}
\def\ece{\end{center}}
\def\nn{\noindent}
\def\nonu{\nonumber}
\def\pbx{\partial_x}


\def\ptl{\partial}
\def\al{\alpha}
\def\be{\beta}
\def\ga{\gamma} 
\def\Ga{\Gamma}
\def\de{\delta} \def\De{\Delta}
\def\ep{\epsilon}
\def\vep{\varepsilon}
\def\ze{\zeta}
\def\et{\eta}
\def\th{\theta} \def\Th{\Theta}
\def\vth{\vartheta}
\def\io{\iota}
\def\ka{\kappa}
\def\la{\lambda} 
\def\La{\Lambda}
\def\rh{\rho}
\def\si{\sigma} \def\Si{\Sigma}
\def\ta{\tau}
\def\up{\upsilon} 
\def\Up{\Upsilon}
\def\ph{\phi} 
\def\Ph{\Phi}
\def\vph{\varphi}
\def\ch{\chi}
\def\ps{\psi} 
\def\Ps{\Psi}
\def\om{\omega} 
\def\Om{\Omega}

\def\lbr{\left(}
\def\rbr{\right)}
\def\half{\frac{1}{2}}
\def\CVO#1#2#3{\!\left( \matrix{ #1 \cr #2 \ #3 \cr} \right)\!}

\def\vol#1{{\bf #1}}
\def\nupha#1{Nucl. Phys. \vol{#1} }
\def\phlta#1{Phys. Lett. \vol{#1} }
\def\phyrv#1{Phys. Rev. \vol{#1} }
\def\PRL#1{Phys. Rev. Lett \vol{#1} }
\def\prs#1{Proc. Roc. Soc. \vol{#1} }
\def\PTP#1{Prog. Theo. Phys. \vol{#1} }
\def\SJNP#1{Sov. J. Nucl. Phys. \vol{#1} }
\def\TMP#1{Theor. Math. Phys. \vol{#1} }
\def\ANNPHY#1{Annals of Phys. \vol{#1} }
\def\PNAS#1{Proc. Natl. Acad. Sci. USA \vol{#1} }

For last several years important progress has been achieved in 
understanding the structure of $N=2$ supersymmetric gauge 
theories\cite{SW,SW2}. The low energy description of these theories
can be encoded by Riemann surfaces or complex algebraic curves and 
the integrals of meromorphic one differentials over the periods of them.
Exact effective actions of these theories 
can be described by holomorphic functions,
so-called prepotentials.
There are many works which connect these low-energy effective
theories with known integrable systems. 
Let us recall some known facts which are relevant to the problems
we address in the present paper. 
To relate effective theories with integrable systems, one needs 
averaging over fast oscillations, i.e. Whitham averaging.
It has been analyzed in Ref.\cite{GKMMM} 
that the periods of the modulated Whitham solution
of periodic Toda lattice give rise to the mass
spectrum in the BPS saturated states.
Furthermore in Ref.\cite{MW} this framework of Whitham 
dynamics for the Toda lattice
was generalized to other gauge groups. 
For the case of $SU(N_{c})$ gauge theory with a single
hypermultiplet in the {\it adjoint} representation, 
the corresponding integrable system was found in Ref.\cite{DW} 
and was recognized to be the elliptic spin model of Calogero\cite{Mar}. 
This connection has been developed in Ref.\cite{IM} by identifying the 
coupling constant of Calogero system with the mass of a 
hypermultiplet in the adjoint
representation, starting from the Lax operator for the Calogero model and
calculating the full spectral curve explicitly.
What was lacking so far was the integrable system related to 
gauge theories coupled to {\it massive } hypermultiplets 
in the {\it fundamental } representation. 

In this letter, we will consider $N =2$ supersymmetric $SU(N_c)$ gauge
theories with $N_c$ colors and $N_f$ flavors. The field content of the 
theories
consists, in terms of $N=1$ superfields, a vector multiplet $W_\al$,
a chiral multiplet $\Ph$, and two chiral superfields $Q^i_a$ and
$\tilde{Q}_{ia}$ where $i= 1, \cdots, N_f$ and $a =1, \cdots, N_c$.
The superpotential reads,
\beq
 W = \sqrt{2} \tilde{Q}_i \Ph Q ^i + \sum_{i=1}^{N_f} m_i \tilde{Q}_i Q^i,
\eeq
where $m_i$'s are the bare quark masses and color indices are suppressed.
The curve representing the moduli space with
$N_{f} < N_{c}$ case is as follows\cite{HO}:
\bea
y^2=(x^{N_{c}}-\sum_{i=2}^{N_{c}} u_{i} x^{N_{c}-i})^2-
\Lambda_{N_{f}}^{2N_{c}-N_{f}} \prod_{i=1}^{N_{f}} (x+m_{i}),
\label{eq:curve}
\eea
where the moduli $u_{i}$'s are the vacuum expectation 
values of a scalar field
of the $N=2$ chiral multiplet, and $m_i$'s are the bare quark masses. 
It turns out that from the point of view of integrable theory, 
$u_{i}$'s correspond to the integrals of motion. 
The second term in Eq.(\ref{eq:curve}) is due to the instanton corrections.
For the $N_c \leq N_f < 2 N_c$ case, 
the correction due to matter is different and the curve
is given as follows\cite{HO}:
\bea
y^2 & = & \left(x^{N_{c}}-\sum_{i=2}^{N_{c}} u_{i} x^{N_{c}-i}
 + \frac{\La^{2N_c-N_f}_{N_f}}{4}\sum_{i=0}^{N_f-N_c} x^{N_f -N_c - i}
 \sum_{j_1 < \cdots <j_i} m_{j_1} \cdots m_{j_i} \right)^2 \nonu \\
& & -
\Lambda_{N_{f}}^{2N_{c}-N_{f}} \prod_{i=1}^{N_{f}} (x+m_{i}).
\label{eq:curve3}
\eea
Clearly the case of $N_f = 0$ corresponds to the periodic 
Toda lattice with $N_c$-particles,
after an appropriate rescaling of the variables\cite{GKMMM}:
\beq
y^2 = P_{N_c} (x)^2 - 1,
\eeq
where $P_n(x)$ is a polynomial of order $n$ whose coefficients  
are Schur polynomials of the Toda lattice.
$N_f \neq 0$ cases are described by the following type of curves
from Eq.(\ref{eq:curve}):
\beq
y^2 = P_n(x)^2 - Q_m (x),
\eeq
where $Q_m(x)$ is a polynomial of order $m(=N_f)$.
It is natural to ask which integrable theories have such spectral curves.
Our motivation is to identify the description of gauge theories with
the data from the integrable theory on the line of 
Refs.\cite{GKMMM,MW,DW,Mar,IM}. 
We will start with the known case of $y^2 = P_3 (x)^2 - a x^2$ (a is
some constant) which 
corresponds to the so called Goryachev-Chaplygin (GC) top, first introduced
by Goryachev\cite{Gor} and later integrated by Chaplygin\cite{Chap}
in terms of hyperelliptic integrals. It has been noted in Ref.\cite{Marsha}
that there exists such a connection.

Let us review the classical mechanics of rotation of a heavy rigid body 
around a fixed point, which are described by the following Hamiltonian:
\bea
H(M,p)=\frac{M_{1}^2}{2I_{1}} +\frac{M_{2}^2}{2I_{2}}+
\frac{M_{3}^2}{2I_{3}}+\gamma_{1} p_{1}+\gamma_{2} p_{2}+
\gamma_{3} p_{3}.
\label{eq:hamil}
\eea
The phase space of this system is six dimensional
: 
$M_i$'s are the components of the angular momentum and $p_i$ are
the linear momenta.
The Poisson brackets of these variables are given by
\beq
\{M_i,   M_j \} = \ep_{ijk} M_k, \;\;\; \{M_i, p_j \}= \ep_{ijk} p_k,   
\;\;\; \{ p_i, p_j \} =0, \;\;\; i,j,k=1,2,3.     
\eeq
where $\ep_{ijk}$ is an antisymmetric tensor.
$I_{1}, I_{2}, I_{3}$ are the principal
moments of inertia of the body and $\gamma_{1}, \gamma_{2}, \gamma_{3}$
are the coordinates of the center of mass. 
There are four known integrable cases for the Hamiltonian in 
Eq.(\ref{eq:hamil}). 
There is always one obvious integral of motion, the energy. 
It is necessary to get one extra integral independent of 
the energy for complete integrability 
according to Liouville's theorem\cite{DKN}.

1) Euler's case  (1750): $\gamma_{1}=\gamma_{2}=\gamma_{3}=0$. 
The extra integral
is $M_{1}^2+M_{2}^2+M_{3}^2$. The symmetry group is
$SO(3)$.

2) Lagrange's case (1788): $I_{1}=I_{2}, \gamma_{1}=\gamma_{2}=0$. 
The new integral
is $M_{3}$. The corresponding symmetry group is $SO(3,1)$. 

3) Kowalewski's case (1889): $ I_{1}=I_{2}=2I_{3}, \gamma_{3}=0$. 
The extra integral can be found by the Painlev\'{e} test or the
Kowalewski's asymptotic
method. Here the symmetry group is $SO(3,2)$.       

4) Goryachev-Chaplygin's case (1900): 
$I_{1}=I_{2}=4I_{3}, \gamma_{3}=0$ and 
$M_{1} p_{1}+M_{2} p_{2}+M_{3} p_{3}=0$ which leads to a new integral of 
motion.  
This is a system which can be integrated in terms of hyperelliptic integrals
after separation of the variables.
In doing so, we obtain the following curve, 
\beq
y^2 = 4\mu ^2 x^2 -\left(x^3- H x -4 G \right)^2,
\label{eq:curvegc}
\eeq
where $\mu$ is a parameter,
$H$ and $G$ are the Hamiltonian and the GC integral\cite{Koz}.
The equations of motion of the GC top with some rescalings are as follows :
\bea      
& & \dot{M_{1}} = 3 M_{2} M_{3}, \;\;\;\;\; 
\dot{M_{2}} = -3 M_{1} M_{3}-2 p_{3},
\;\;\;\;\; \dot{M_{3}} =2 p_{2}, \nonu \\
& & \dot{p_{1}} = 4 M_{3} p_{2}- M_{2} p_{3}, \;\;\;\;\; 
\dot{p_{2}}=M_{1} p_{3}-
4 M_{3} p_{1}, \;\;\;\;\; \dot{p_{3}} = M_{2} p_{1}-M_{1} p_{2},
\label{eq:diffgc}
\eea
where $\cdot$ means the time derivative.
The curve in Eq.(\ref{eq:curvegc}) 
can also be obtained from the Lax operator for the GC top\cite{BK}.
It turns out that 
the Lax operator is related to that of the Kowalewski top in the $4 \times 4$
spinor representation of $SO(3,2)\simeq Sp(4)$. 
Although the underlying structure is not fully 
understood yet, the Lax operator for the GC top is then 
obtained by removing the first row and column
of the Lax operator for the Kowalewski top 
and is given as follows\cite{BK}: 
\bea
L(z)=
\left( \begin{array}{ccc}
0 & -i p_3/z & M_2-i M_{1} \\
i p_{3}/{z} & 2 i M_{3} & -2 i z +(p_{2}-i p_{1})/z  \\
-M_{2}-i M_{1} & 2 i z+(p_{2}+ i p_{1})/z & -2 i M_{3}  \\
\end{array} \right).
\eea
This Lax operator depends on the phase space variables, $M_i, p_i$ 
and on the spectral parameter, $z$.
Then we can show that the Lax equation $\dot{L}=[ L, A ]$ gives 
the equations of motion with the following matrix $A$: 
\bea
A(z)=
\left( \begin{array}{ccc}
3 i M_{3} & 0 & M_{2}-i M_{1} \\
0 & 2 i M_{3} & -2 i z   \\
-M_{2}-i M_{1} & 2 i z & -2 i M_{3}  \\
\end{array} \right).
\eea
Now it is easy to calculate the spectral curve from the equation
${\bf C}:{\rm Det}( L(z)-x I ) =0$, which gives the spectral curve
as follows:
\bea
x^3+2 x H -2 i G-x (4 z^2+\frac{\lambda^2}{z^2}) =0,
\eea
where $H=\frac{1}{2} ( M_{1}^2+M_{2}^2+4 M_{3}^2 )-2 p_{1}$ is the
Hamiltonian, and $G=M_{3} (M_{1}^2+M_{2}^2)+2 M_{1} p_{3}$ is the 
GC integral. 
We also have the following constraints: 
\beq
p_{1}^2+p_{2}^2+p_{3}^2=\lambda^2,\;\;\; {\rm and}\;\;\; M_{1} p_{1}+
M_{2} p_{2}+M_{3} p_{3}=0 . 
\eeq
Now we see that the spectral curve
depends on the special combinations of $M_i, p_i$'s, which are nothing but 
the integrals of motion.
By introducing $y=x (4 z^2-\frac{\lambda^2}{z^2})$, we thus get
\bea
y^2=(x^3+2 H x-2 i G)^2-16 \lambda^2 x^2,
\label{eq:curve1}
\eea
which are the same as Eq.(\ref{eq:curvegc}) with some rescalings.
To relate this to the curve of supersymmetric gauge
theory we make the following
substitutions:
\bea 
H \rightarrow -\frac{1}{2} u_{2}, \;\;\;\;\; 
G \rightarrow -\frac{i}{2} u_{3},
\;\;\;\;\; \lambda^2 \rightarrow \frac{1}{16} \Lambda_{2}^{4},
\eea
and it is easy to see that Eq.(\ref{eq:curve1}) exactly coincides 
with Eq.(\ref{eq:curve}) for the particular
case of $N_{c}=3, N_{f}=2$ and $m_{1}=m_{2}=0$.

Since we have seen the intimate relation between the GC top and 
the supersymmetric $SU(3)$ gauge theory with two flavor
massless hypermultiplets, it is natural for us to extend
this to the massive case. For this purpose we need an integrable system
which has both the GC top and the three body 
Toda lattice as particular limits, because the latter corresponds to
pure gauge theory with no matter.
The Hamiltonian system which realizes this is hard to imagine,
but there exists a system of coupled seven nonlinear differential
equations in mathematical literature\cite{BvM}.
This system has the following ``equations of motion":
\bea
\dot{z_{1}} & = & -8 z_{7}, \ \ \ \ \ \ 
\dot{z_{2}}  =  4 z_{5}, \nonu \\
\dot{z_{3}} & = & 2 (z_{4} z_{7}-z_{5} z_{6}), \ \ \ \ \ \ 
\dot{z_{4}}  =  4 z_{2} z_{5}-z_{7}, \nonu \\
\dot{z_{5}} & = & z_{6}-4 z_{2} z_{4}, \ \ \ \ \ \ 
\dot{z_{6}}  =  -z_{1} z_{5}+2 z_{2} z_{7}, \nonu \\
\dot{z_{7}} & = & z_{1} z_{4}-2 z_{2} z_{6}-4 z_{3},
\label{eq:diff}
\eea
and the following five constants of motion:
\bea
6 a & = & z_{1}+4 z_{2}^2-8 z_{4}, \nonu \\
2 b & = & z_{1} z_{2}+4 z_{6}, \nonu \\
c   & = & z_{4}^2+z_{5}^2+z_{3}, \nonu \\
d   & = & z_{4} z_{6}+ z_{5} z_{7}+z_{2} z_{3}, \nonu \\
e   & = & z_{6}^2+z_{7}^2- z_{1} z_{3}.
\label{eq:integral}
\eea

Although the Lax operator for this system is not readily available, 
we can still apply the asymptotic method due to Kowalewski to this system 
and take $z_i=t^{-n_i} \sum_{j=0}^{\infty} A^{i}_{j} t^{j}$ where
$n_i$'s are positive integers\cite{SE,BvM}. 
Substituting these Laurent expansions
into the system of Eqs.(\ref{eq:diff}) and (\ref{eq:integral}),
one finds $n_i=1$ for $i=1, 2, 3$,
$n_i=2$ for $i=4, 5, 6, 7$ and a relation between the coefficients of 
$A^{i}_{j}$'s.
Then we obtain
the Laurent solutions for this system with seven parameters, 
five of which are from the constants of motion, $a, b, c, d, e$
and two additional ones $x$ and $y$ where they satisfy the equation
for an hyperelliptic curve\cite{BvM}:
\bea
y^2=P(x)^2 -4 Q(x).
\label{eq:curve2}
\eea
Here the  
polynomials of $x$ are
$P(x) = 2 x^3-3 a x+b $ and $ Q (x) = 4 c x^2+4 d x+ e$.
We clearly see that with the following  substitution this gives the algebraic
curves given in Eq.(\ref{eq:curve}) of $N=2$ supersymmetric  $SU(3)$  
gauge theories with massive quarks of two flavors of masses $m_1$ and $m_2$:
\bea
& & y \rightarrow 2 y, \;\;\;\;\; a \rightarrow \frac{2}{3} u_{2}, 
\;\;\;\;\;
b \rightarrow -2 u_{3}, \nonu \\
& & c \rightarrow \frac{1}{4} \Lambda_{2}^{4}, \;\;\;\;\; d \rightarrow
\frac{\Lambda_{2}^{4}}{4} (m_{1}+m_{2}), \;\;\;\;\;
e \rightarrow \Lambda_{2}^{4} m_{1} m_{2}.
\label{eq:eq19}
\eea
When we consider the case of $c=0$, then this
leads to gauge theory coupled to one massive quark of 
mass $m_1$ or massless one$(N_f=1)$
after the substitution: 
\bea
& & y \rightarrow 2 y, \;\;\;\;\; a \rightarrow \frac{2}{3} u_{2}, 
\;\;\;\;\;
b \rightarrow -2 u_{3}, \nonu \\
& & d \rightarrow \frac{\Lambda_{1}^{5}}{4} , \;\;\;\;\;
e \rightarrow \Lambda_{1}^{5} m_{1}. 
\eea

For the case of $c=d=0$,
the following transformations
gives us to the usual periodic Toda lattice with three particles:
\bea
& & z_{1}=-e^{q_{1}-q_{2}}+p_{1} p_{2}, \;\;\;\;\;
z_{2}=-\frac{i}{2} p_{3}, \nonu \\
& & z_{3}=-\frac{1}{16} e^{q_{2}-q_{1}}, \;\;\;\;\;
z_{4}=\frac{1}{8} \left( e^{q_{2}-q_{3}}+ 
e^{q_{3}-q_{1}} \right),\nonu \\
& & z_{5}=-\frac{i}{8} \left( e^{q_{2}-q_{3}}-e^{q_{3}-q_{1}} 
\right),\;\;\;\;\;
z_{6}=\frac{i}{8} \left( p_{2} e^{q_{3}-q_{1}} +p_{1} e^{q_{2}-q_{3}} 
\right), \nonu \\ 
& & z_{7}=\frac{1}{8} \left( p_{2} e^{q_{3}-q_{1}}-p_{1}
e^{q_{2}-q_{3}} \right).
\label{eq:tran}
\eea
Here the $q_{i}$'s are the coordinates of the Toda particles 
and $p_{i}$'s are corresponding
momenta. Using Eq.(\ref{eq:tran}),  it is easy to see that the remaining
integrals 
of motion can be written as those of the periodic Toda lattice:
\bea
 -6 a & = & \frac{1}{2} \sum_{i=1}^{3} 
p_{i}^2+\sum_{i=1}^{3} e^{q_{i}-q_{i+1}},
\nonu \\
2 b & = &  -\frac{i}{2} \left( p_{1} p_{2} p_{3} -p_{1} e^{q_{2}-q_{3}}-
p_{2} e^{q_{3}-q_{1}}-p_{3} e^{q_{1}-q_{2}} \right), \nonu \\
 e & = & -\frac{1}{16}
\eea
where $a$ and $b$ are now proportional to the Hamiltonians of Toda system.

It can be easily checked that this extended integrable system reduces to
the GC top when we take $d=e=0$, and this is 
achieved by the following transformations:
\bea
& & z_{1}=M_{1}^2+M_{2}^2, \;\;\;\;\; z_{2}=M_{3}, \;\;\;\;\; 
z_{3}=\frac{1}{4}
p_{3}^2, \nonu \\
& & z_{4}=\frac{1}{2} p_{1}, \;\;\;\;\; z_{5}=\frac{1}{2} p_{2}, \;\;\;\;\; 
z_{6}=
\frac{1}{2} M_{1} p_{3}, \;\;\;\;\; z_{7}=\frac{1}{2} M_{2} 
p_{3}. 
\eea
From (\ref{eq:integral}), we can identify $a, b, c$ in terms of $H, G, 
\lambda$ as follows:  
\bea
a=\frac{H}{3}, \;\;\;\;\; b=\frac{G}{2}, \;\;\;\;\;  c=\frac{\lambda^2}{4}.
\eea
Of course, Eq.(\ref{eq:diff}) becomes Eq.(\ref{eq:diffgc}).
This is the $m_1 = m_2 = 0$ case,  as we see from Eq.(\ref{eq:eq19}).
So the system of equations in Eq.(16) has both the Toda lattice
and GC top as particular limits.

There are still several issues to be studied further.
It was already noticed in Ref.\cite{GKMMM} that Whitham solution which
is necessary to produce the effective action for slow variables
plays the important role of nonperturbative analog of renormalization group 
approach of perturbative quantum field theory. 
If one wishes to obtain the
prepotentials which are needed for exact effective action in 
supersymmetric gauge theory, we should consider quasiclassical $\tau$
fuctions in the context of integrable theory as in the case of pure
gauge theory\cite{NT}.
It would be interesting to find out intimate relation between them
by using the explicit form of Baker-Akiezer 
function for GC top\cite{BK}. 
In fact, in Ref.\cite{EY}, the Whitham equations for $SU(N_c)$ 
gauge theory with $N_f$ matter were considered, 
where to make connection with the gauge 
theory, all but one of the ``time" variables was set to zero for the massless
case. This is consistent with the form of the Baker-Akiezer function
of GC top\cite{BK} which depends only on one time variable $t$.
There are algebraic curves for higher rank cases with generic $N_c$ and 
$N_f$, as given by Eqs.(\ref{eq:curve}) and (\ref{eq:curve3}).
The obvious thing to do would be to obtain a `higher' dimensional 
generalization of GC top, at least for the massless cases.
There in fact exists multi-dimensional generalization\cite{BRS} of Kowalewski top,
by using the Lie algebra based on $SO(p,q)$ where the top is in $p$ 
dimensional under $q$ different constant fields. 
However, the spinor representation which was crucial in obtaining the
Lax operator for GC top, is
not available in higher dimensions.
Furthermore, the mysterious fact that removal of the first column and
row in the Lax operator of Kowalewski top is related to GC top is 
yet to be understood for the generalization to the higher dimensional 
cases\footnote{We thank A.I. Bobenko and M.A. Semenov-Tian-Shansky
for correspondences on this matter.}. 
Nevertheless, with all the results from
supersymmetric gauge theories pointing to the existence of higher dimensional
generalizations, it is quite tempting to speculate that there exists 
higher dimensional GC top.
As regards the massive cases, better understanding of the variables,
$z_i$'s$(i=1, \cdots, 7)$ in Eq.(\ref{eq:diff}) are needed as well as
the symmetry of the system. In this respect, the relation to the quadratic
algebra might shed further light in the problem in Refs.\cite{KT,Kul}.
Of course it would be nice to find similar integrable theories for other
gauge theories coupled with real matter. 

This work is supported in part by Ministry of Education (BSRI-95-2442), 
KOSEF-JSPS exchange program, and by  KOSEF 961-0201-001-2. 
The work of S.N. is also supported by CTP/SNU.

\end{document}